\def\apj{\rm ApJ}

\def\aj{\rm AJ}
\def\mnras{\rm MNRAS}

\def\lesssim{\mathrel{\hbox{\rlap{\hbox{\lower4pt\hbox{$\sim$}}}\hbox{$<$}}}}
\def\gtrsim{\mathrel{\hbox{\rlap{\hbox{\lower4pt\hbox{$\sim$}}}\hbox{$>$}}}}

\documentclass{kapproc} 

\usepackage{procps} 

\usepackage[dvips]{graphicx}

\upperandlowercase

\kluwerbib 

\begin{document}



\articletitle[Resolving the Stellar Outskirts of M31 and M33]{Resolving the Stellar Outskirts of M31 and M33}















\author{Annette Ferguson\altaffilmark{1}, Mike Irwin\altaffilmark{2}, Scott Chapman\altaffilmark{3}, 
Rodrigo Ibata\altaffilmark{4}, Geraint Lewis\altaffilmark{5}, Nial Tanvir\altaffilmark{6}}

\affil{\altaffiltext{1}{Institute for Astronomy, University of Edinburgh, UK}
\altaffiltext{2}{Institute of Astronomy, University of Cambridge, UK}
\altaffiltext{3}{California Institute for Technology, Pasadena, USA}
\altaffiltext{3}{Observatoire de Strasbourg, Strasbourg, France}
\altaffiltext{3}{Institute of Astronomy, University of Sydney, Australia}
\altaffiltext{3}{Centre for Astrophysics Research, University of Hertfordshire, UK}}





 \begin{abstract} Many clues about the galaxy assembly process lurk in
the faint outer regions of galaxies.  The low surface brightnesses of
these parts pose a significant challenge for studies of diffuse
light, and few robust constraints on galaxy formation
models have been derived to date from this technique.  Our group has
pioneered the use of extremely wide-area star counts to quantitatively
address the large-scale structure and stellar content of external galaxies at
very faint light levels. We highlight here some results from our
imaging and spectroscopic surveys of M31 and M33.
\end{abstract}


\section{Introduction}

The study of galaxy outskirts has become increasingly important in
recent years. From a theoretical perspective, it has been realised
that many important clues about the galaxy assembly process should lie
buried in these parts. Cosmological simulations of disk galaxy
formation have now been carried out by several groups and have led to
testable predictions for the large-scale structure and stellar content
at large radii -- for example, the abundance and nature of stellar
substructure (e.g.  Bullock \& Johnston 2005, Font et
al. 2005), the ubiquity, structure and content of stellar halos and
thick disks (e.g. Abadi et al. 2005, Governato et al. 2004, Brook et
al. 2005) and the age distribution of stars in the outer regions of thin disks
(e.g.  Abadi et al. 2003).

Bullock \& Johnston (2005) find that Milky Way-like galaxies will have
accreted 100-200 luminous satellites during the last 12~Gyr and that
the signatures of this process should be readily visible at surface
brightnesses of V$\sim 30$ magnitudes per square arcsec and lower. Although
traditional surface photometry at such levels (roughly 9 magnitudes
below sky) remains prohibitive, star count analyses of nearby galaxies
have the potential to reach these effective depths (e.g. Pritchet \&
van den Bergh 1994). The requirement of a large survey area (to
provide a comprehensive view of the galaxy) and moderate-depth imagery
can now be achieved in a relatively straightforward manner using
wide-field imaging cameras attached to medium-sized telescopes.

\section{The INT WFC Surveys of M31 and M33}

In 2000, we began a program to map the outer regions of our nearest
large neighbour, M31, with the Wide-Field Camera equipped to the INT
2.5m. The success of this program led us to extend our survey to M33 in
the fall of 2002. To date, more than 45 and 7 square degrees have been
mapped around these galaxies respectively. Our imagery reaches to
V$\sim$24.5 and {\sl i}$\sim$23.5 and thus probes the top 3 magnitudes
of the red giant branch (RGB) in each system. The raw data are
pipeline-processed in Cambridge and source catalogues are produced
containing positions, magnitudes and shape parameters.  The M31 survey
currently contains more than 7 million sources, and the M33 survey more than 1
million.  Magnitude and colour cuts are applied to point-like
sources in order to isolate distinct stellar populations
and generate surface density maps (see Figure 1). The faint
structures visible by eye in Figure 1 have effective V-band surface
brightnesses in the range 29-30 magnitudes per square arcsec.  Early
versions of our M31 maps have been discussed in Ibata et al. (2001),
Ferguson et al. (2002) and Irwin et al. (2005).

\begin{figure}[t]
{\hspace{-1.5cm}\includegraphics[width=15cm]{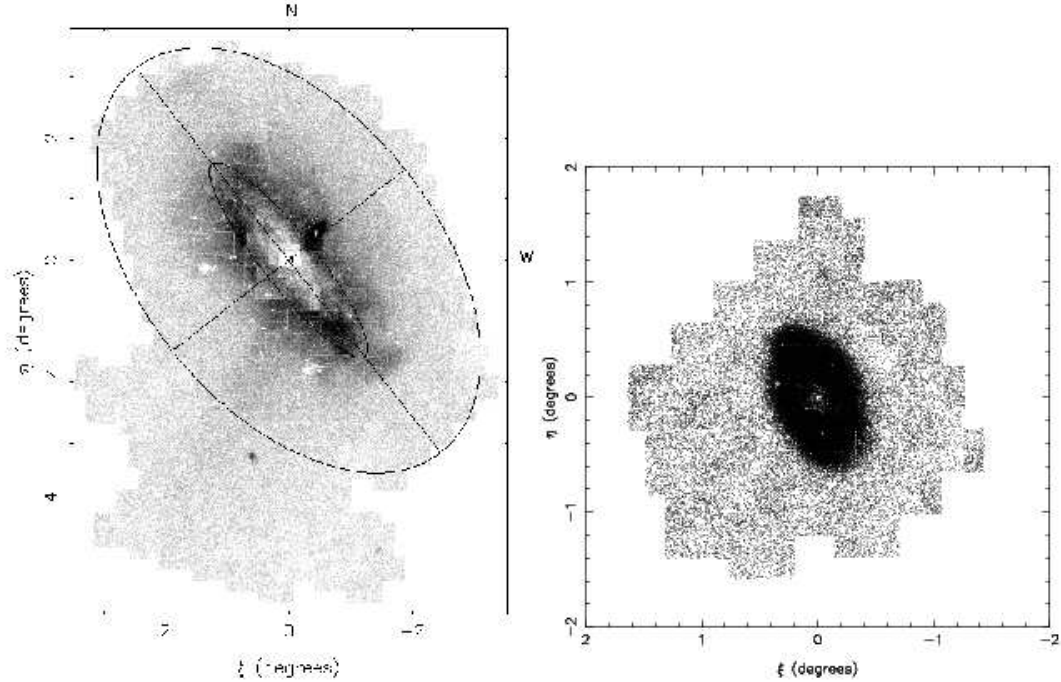}}
\caption{INT/WFC RGB star count maps of M31 (left panel) and M33 (right 
panel). }
\end{figure} 

\subsection{Results for M31}

The left-hand panel of Figure 1 shows the distribution of blue
(i.e. presumably more metal-poor) RGB stars in and around M31.  A
great deal of substructure can be seen including the giant stream in
the south-east, various overdensities near both ends of the major axes, a
diffuse extended structure in the north-east and a loop of stars projected near
NGC~205.  

{\sl Origin of the Substructure:} Do the substructures in M31
represent debris from one or more satellite accretions, or are they
simply the result of a warped and/or disturbed outer disk?  We are
addressing these issues with deep ground-based imagery from
the INT and CFHT, Keck-10m spectroscopy and deep HST/ACS
colour-magnitude diagrams (CMDs). Our findings to date can be
summarized as follows:

\begin{itemize}
\item{M31 has at least 12 satellites lying within a
    projected radius of 200~kpc.  The bulk of these systems,
    the low-luminosity dwarf spheroidals, are unlikely to be associated with the
    stellar overdensities since their RGB stars are much bluer than
    those of the substructure (Ferguson et al. 2002).}
  
\item{The combination of line-of-sight distances and radial velocities
    for stars at various locations along the giant stellar stream
    constrains the progenitor orbit (e.g.  McConnachie et al. 2003,
    Ibata et al. 2004). Currently-favoured orbits do not connect the
    more luminous inner satellites (e.g. M32, NGC~205) to the stream
    in any simple way however this finding leaves some remarkable
    coincidences (e.g. the projected alignment on the sky, similar
    metallicities) yet unexplained.}
  
\item{Deep HST/ACS CMDs reaching well below the horizontal branch
    reveal different morphologies between most substructures in
    the outskirts of M31 (Ferguson et al. 2005, see Figure 2).  These
    variations reflect differences in the mean age and/or metallicity
    of the constituent stellar populations. Analysis is underway to
    determine whether multiple satellite accretions are required, or
    whether consistency can be attained with a single object which has
    experienced bursts of star formation as it has orbited M31.  The
    giant stream is linked to another stellar overdensity, the NE
    shelf, on the basis of nearly identical CMD morphologies and RGB
    luminosity functions; indeed, this coupling seems likely in view
    of progenitor orbit calculations (e.g. Ibata et al.
    2004).}
\end{itemize}

\begin{figure}[t]
{\hspace{-1cm}\includegraphics[width=10cm,angle=90]{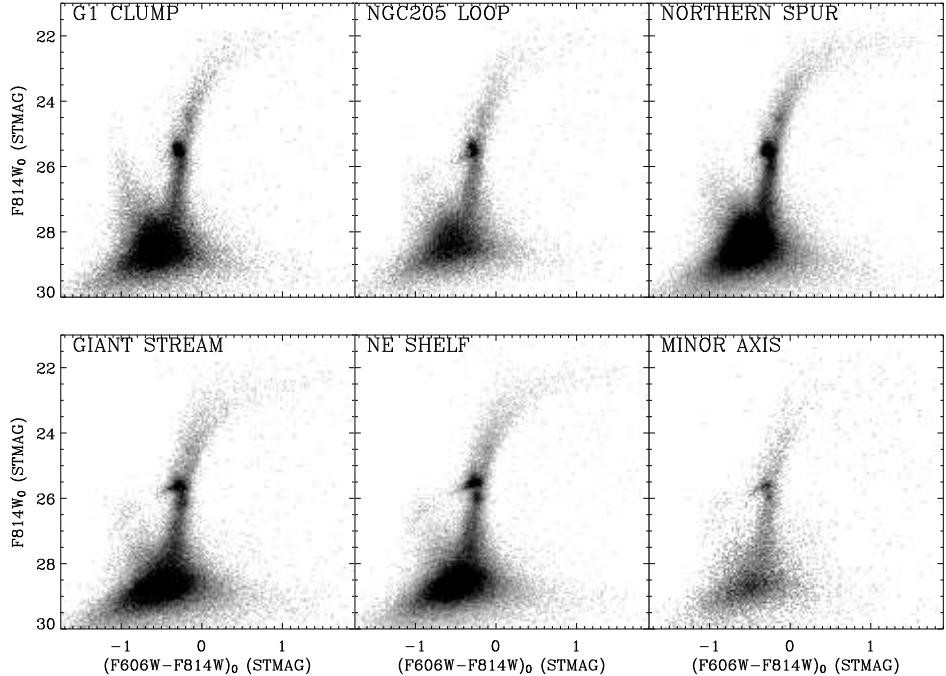}}
\caption{HST/ACS Hess diagrams of six halo fields in M31.  Clear differences are
apparent between most CMDs, indicating genuine stellar population 
variations between the substructures.}
\end{figure}   

{\sl Smooth Structure:} The INT/WFC survey provides the first
opportunity to investigate the smooth underlying structure of M31 to
unprecedented surface brightnesses. We have used the dataset to
map the minor axis profile from the innermost regions to $\gtrsim 55$~kpc 
(Irwin et al.  2005). Figure 3 shows how the combination of
inner diffuse light photometry and outer star count data can be used to
trace the effective {\sl i}-band surface brightness profile 
to $\sim 30$~magnitudes per square arcsec. The profile
shows an unexpected flattening (relative to the inner R$^{1/4}$ decline)
at large radius, consistent with the presence of an additional shallow power-law
stellar component (index $\approx -2.3$) in these parts.  This component may
extend out as far as 150~kpc (Guhathakurta, these proceedings).  Taken
together with our knowledge of the Milky Way halo, this finding
supports the ubiquitous presence of power-law stellar halos around 
bright disk galaxies (see also Zibetti et al. 2004, Zibetti \& Ferguson 2004).

\begin{figure}
{\includegraphics[width=6cm]{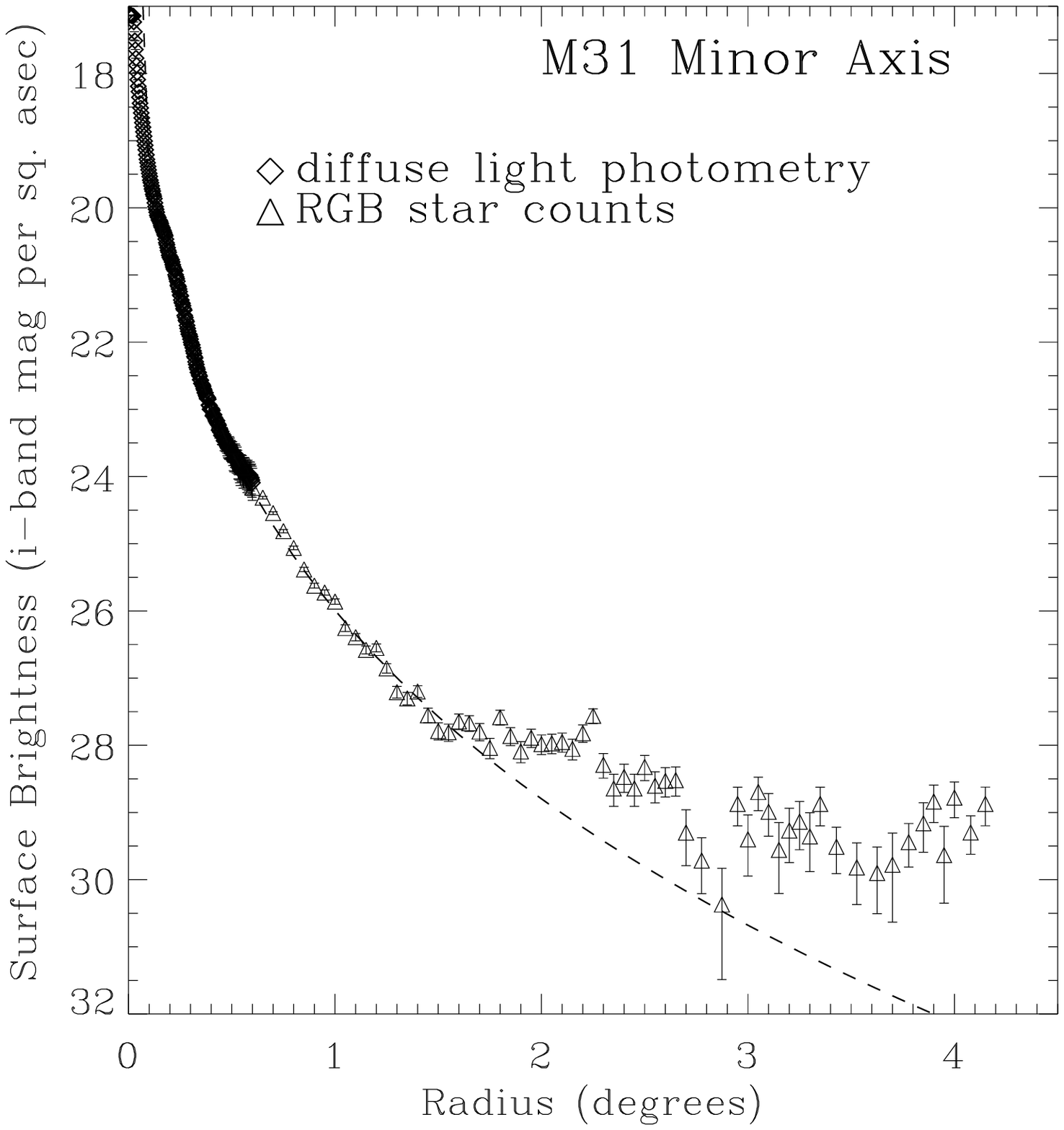}\includegraphics[width=6cm]{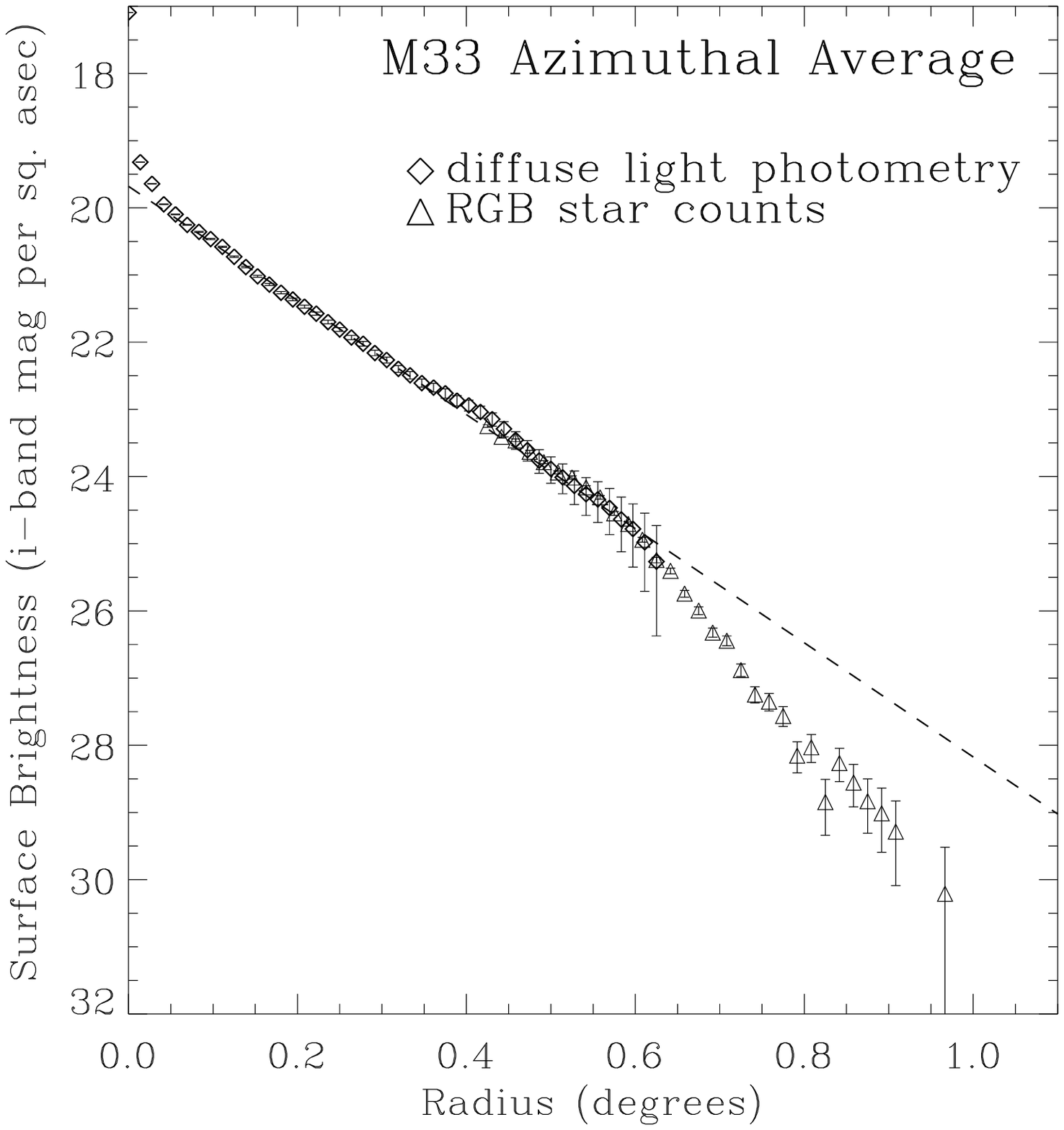}}
\caption{Effective {\sl i}-band surface brightness profiles for M31 (left)
and M33 (right). The inner points are
derived from surface photometry and the outer ones from RGB star counts.
Overplotted are a de Vaucouleurs R$^{1/4}$ law with r$_e=0.1$ degrees (left
panel) and an exponential profile with r$_0=0.1$ degrees (right panel). }
\end{figure} 

The kinematics of M31's outer regions are being probed with Keck
DEIMOS spectroscopy (e.g. Ibata et al. 2005).  Two surprising results
have emerged from our program so far. Firstly, there is a high degree
of rotational support at large radius, extending well beyond the
extent of M31's bright optical disk. Secondly, the overall coherence
of this kinematic component is in striking contrast to its clumpy substructured
appearance in the star count maps.  Further work is underway 
to understand the nature and origin of this rotating component.

\subsection{Results for M33}

The right-hand panel of Figure 1 shows the RGB map of the low mass
system, M33.  Although it has the same limiting absolute depth as the
M31 map, the stellar density distribution is extremely smooth and
regular (Ferguson et al. 2006, in preparation).  To a limiting depth
of $\sim 30$~magnitudes per square arcsec (readily visible by eye
here), the outer regions of M33 display no evidence for stellar
substructure (c.f.  the simulations of Bullock \& Johnston 2005).
Equally surprising, our analysis of the isophote shape as a function of
radius indicates no evidence for any twisting or asymmetries. 
M33 appears to be a galaxy which has evolved in relative
isolation.

The radial {\sl i}-band profile of M33 has been quantified via azimuthally-averaged
photometry in elliptical annuli of fixed PA and inclination (Figure 3). 
The inner parts of the profile are constructed
from diffuse light photometry, whereas the outer regions are derived
from RGB star counts.  The luminosity profile displays an exponential
decline out to $\sim 8$~kpc (roughly 4.5 scalelengths) beyond which it
significantly steepens. This behaviour is reminiscent of the ``disk
truncations'' first pointed out by van der Kruit in the 80's, but until
now not seen directly with resolved star counts.  The steep
outer component dominates the M33 radial light profile out to at least 14~kpc
and limits the contribution of any shallow power-law stellar halo
component in M33 to be no more than a few percent of the disk luminosity
(Ferguson et al. 2006, in preparation).

\section{Future Work}

Quantitative study of the faint outskirts of galaxies provides
important insight into the galaxy assembly process. The outskirts
of our nearest spiral galaxies, M31 and M33, exhibit intriguing
differences in their large-scale structure and stellar content.  While
M31 appears to have formed in the expected hierarchical fashion, M33
shows no obvious signatures of recent accretions. Observations of
the outer regions of additional galaxies are required to determine
which of these behaviours is most typical of the general disk
population.




%



\end{document}